\documentclass[manuscript,screen]{acmart}
\AtBeginDocument{%
  }

\begin{document}

\title{Hybrid Drawing Solutions in AR: Bitmap-to-Vector Techniques on 3D Surfaces}

\author{Pengcheng Ding}
\email{ding.pen@northeastern.edu}
\author{Yedian Cheng}
\email{cheng.yed@northeastern.edu}
\author{Mirjana Prpa}
\email{m.prpa@northeastern.edu}
\affiliation{%
  \institution{Northeastern University}
  \city{Vancouver}
  \state{British Columbia}
  \country{Canada}
}

\begin{abstract}
Recent advancements in augmented reality (AR) and virtual reality (VR) have significantly enhanced workflows for drawing 3D objects. Despite these technological strides, existing AR tools often lack the necessary precision and struggle to maintain quality when scaled, posing challenges for larger-scale drawing tasks. This paper introduces a novel AR tool that uniquely integrates bitmap drawing and vectorization techniques. This integration allows engineers to perform rapid, real-time drawings directly on 3D models, with the capability to vectorize the data for scalable accuracy and editable points, ensuring no loss in fidelity when modifying or resizing the drawings.
We conducted user studies involving professional engineers, designers, and contractors to evaluate the tool’s integration into existing workflows, its usability, and its impact on project outcomes. The results demonstrate that our enhancements significantly improve the efficiency of drawing processes.Specifically, the ability to perform quick, editable, and scalable drawings directly on 3D models not only enhances productivity but also ensures adaptability across various project sizes and complexities.
\end{abstract}




\maketitle

\section{Introduction}

Visualizing complex objects in two dimensions is limiting, as it restricts our ability to perceive accurate spatial relationships and dimensions. Immersive technologies such as Virtual Reality (VR) and Augmented Reality (AR) are increasingly being integrated into professional workflows to address these limitations. These technologies enhance our ability to visualize 3D objects with greater detail and precision, providing experiences that mimic interacting with physical objects but without the constraints of physical space.

Currently, VR and AR technologies are becoming significant in fields such as engineering. Advancements in photogrammetry\cite{Gomarasca2009Elements, Muraki2001A} have enabled designers to directly create three-dimensional models from photographs, transforming traditional two-dimensional workstations into immersive three-dimensional environments. Augmented Reality, in particular, facilitates the projection of 3D models into the real world, providing a novel mode of interactive visualization that effectively merges virtual objects with the physical environment. This capability significantly enhances the engineering process by allowing professionals to view and manipulate 3D models in real-time\cite{Dong2013Collaborative}, thereby ensuring more precise outcomes. Consequently, there is a growing demand in the technical community for capabilities such as direct drawing on 3D models in augmented reality, highlighting the need for tools that can perform precise annotations and drawing on 3D objects to enhance detail-oriented tasks.

Research in AR and VR sketching has developed over time, and a wealth of valuable research already exists, allowing us to delve deeper into immersive sketching with these technologies.  Early systems like HoloSketch\cite{Deering95} pioneered 3D modeling but struggled with more structured and detailed creation . Technologies such as Tilt Brush\cite{tan2022examining} improved control but still lack functionalities like stroke capture\cite{Jackson16,Schkolne01,Israel13,Israel09,Arora17}. Efforts like FreeDrawer\cite{Wesche01} and haptic feedback integration\cite{Kamuro11,Keefe07} aim to enhance precision. Arora et al.\cite{Arora17} highlighted the challenges of accurately creating curves on virtual surfaces in VR, even on simple planar surfaces. Furthermore, hybridizing 2D and 3D sketching, as seen in PintAR\cite{Gasques19,Gasques2019} and TabletInVR\cite{Surale19}, improves 3D surface drawing accuracy. SymbiosisSketch\cite{Arora18} and VRSketchIn\cite{Drey20} combine air-pen and tablet inputs, but managing complex surfaces like spheres remains difficult. Advanced projection techniques, used in SecondSkin\cite{DePaoli15}, Skippy\cite{Krs17}, and CASSIE\cite{Yu2021}, align drawn trajectories to object surfaces, improving mid-air drawing accuracy\cite{Arora21}.Using the vector method to draw in has also made some progress. A key study called VTracer\cite{Sanford20} is a tool that allows for the transformation of raster images into vector graphics. By leveraging this frame field, the algorithm creates vectorizations that align closely with the original raster images. These researches are priceless resources for us to start our research.

Even with these improvements, modern AR technologies frequently fail to satisfy the particular requirements of engineers. No drawing technique currently in use can meet both the needs of quick sketching and preserving accuracy and edibility in large-scale models.

To overcome the shortcomings of existing augmented reality technologies, our approach builds a custom AR framework for engineering applications by combining the immersive features of the Microsoft HoloLens 2\cite{hololens} with an advanced hybrid drawing tool built on Unity. This drawing tool uniquely combines the advantages of rapid sketching with the scalability and 
 editability of vector graphics.

In summary, our contributions are as follows:

\begin{itemize} 
\item We have developed a tool for rapid drawing on 3D model surfaces in AR. This innovative tool allows for initial bitmap sketches to be made directly within the AR environment, facilitating immediate and intuitive interaction with 3D models. 
\item To preserve clarity, editability, and scale, we enable the vectorization of sketches. After the initial sketching, our tool can vectorize these sketches, ensuring that the drawings maintain their editable, detail, and scale. 
\item We evaluated the tool's usability and efficiency in real-world circumstances. This study contributed to uncover practical issues and user preferences, which are critical for iterative development.
\end{itemize}

This novel framework revolutionizes the conventional drawing process by facilitating a smooth transition from bitmap to vector. This conversion enhances the utility of drawings throughout various phases of the engineering design process, making rapid prototyping and modifications more manageable. Moreover, our hybrid drawing approach ensures that drawings maintain accuracy and precision even at large scales, which is particularly vital for large-scale engineering projects where precision and detail are critical.


\section{RELATED WORK}
Our work builds upon a substantial body of prior research in the areas of immersive sketching and modeling, integration of 2D and 3D sketching techniques, enhancing 3D modeling through projection techniques, and integrating bitmap drawing and vectorization process for enhanced precision in 3D AR Environments. This diverse foundation informs our approach to developing advanced AR tools for engineering applications.

\subsection{Immersive sketching and modeling:}

Immersive sketching and modeling have rapidly evolved through advancements in VR and AR technologies. Introduced in 1995, HoloSketch\cite{Deering95} utilized a 6-degree-of-freedom wand for creating basic 3D shapes, setting a foundational stage for interactive 3D modeling.Despite its innovation, early systems like HoloSketch often suffered from poor motor control, leading to imprecise and noisy mid-air drawings.Subsequent systems have further developed these initial ideas, delving into the creation of freeform 3D curves and swept surfaces\cite{Jackson16,Schkolne01}.Technologies like Google's Tilt Brush\cite{TiltBrush20} have built on these foundations, offering enhanced control and the ability to transform direct 3D inputs into detailed creative outputs. However, these systems still struggle with the inherent inaccuracies of 3D sketching, especially when detailed, structured creation is needed. Current immersive tools lack functionalities like stroke capture and real-time feedback, visual depth cues, and motion parallax , which are standard in traditional 2D sketching\cite{Israel13,Israel09,Arora17}. Efforts like FreeDrawer\cite{Wesche01} have tried to improve precision by regularizing 3D curves to existing geometries, and some have integrated haptic feedback to refine user interaction\cite{Kamuro11,Keefe07}. 
Arora et al.\cite{Arora17} highlighted the challenges of accurately creating curves on virtual surfaces in VR, even on simple planar surfaces. This finding underscores the need for our research into techniques that project 3D strokes onto surfaces.

\subsection{Integration of 2D and 3D sketching techniques:}

Enhancing the precision of 3D surface drawings can be significantly advanced through the innovative hybridization of 2D and 3D sketching techniques.Gasques et al.\cite{Gasques19,Gasques2019} identified the need for this hybrid sketching approach in their PintAR system, which provides a pen and tablet interface without mid-air interactions, focusing on a more grounded drawing experience. Similarly, TabletInVR \cite{Surale19}is a recent initiative that creates a design space for VR 3D solid modeling based on tablet inputs, demonstrating the practical application of tablet-based controls in immersive environments. SymbiosisSketch\cite{Arora18} combines freeform air-pen sketching in AR with pen-and-tablet inputs, facilitating sketches on both flat and curved surfaces within AR, thereby enhancing interaction with real-world objects. However, a significant limitation emerges when the scene exceeds a basic canvas, such as when dealing with continuous surfaces like spheres. This complexity is highlighted by the challenges of extending beyond traditional surface patch topology to accommodate surfaces with multiple boundaries or closed forms like spheres. VRSketchIn\cite{Drey20} further explores this concept within its design space, categorizing prior art and defining metaphor groups of sketching interactions for VR. This system was implemented and evaluated through usability walkthrough with participants, emphasizing its primary application in artistic creation rather than technical tasks. The findings suggest that while 2D and 3D integration is beneficial for artistic endeavors, it remains challenging to manage on larger and continuous surfaces like spheres.

\subsection{Enhancing 3D Model Drawing through Projection Techniques}
Advanced projection techniques in 3D modeling play a crucial role in enhancing the quality of digital representations by mapping intricate contours onto object surfaces. Techniques such as SecondSkin\cite{DePaoli15} and Skippy\cite{Krs17} utilize insights into the spatial relationships between drawn strokes and the 3D object, inferring 3D curves that accurately follow the object's surface. In these models, drawn trajectories can be aligned to predefined lines\cite{Yu21} or planes \cite{Machuca18}, or integrated into 3D curve networks\cite{Yu2021}. Furthermore, CASSIE\cite{Yu2021} has developed a 3D optimization approach that enables the automatic connection and predictive smoothing of curves, enhancing the user's ability to create more coherent 3D structures. Graffiti-style painting with a spray can is widely recognized as a standard practice in commercial immersive paint and sculpt software like Medium\cite{MediumAdobe2021} and Gravity Sketch\cite{GravitySketch2020}. This method often employs a closest-point projection that simulates drawing on the 3D object without direct physical contact, similar to the method used by the "guides" tool in Tilt Brush\cite{TiltBrush20}. While existing approaches handle each mid-air point separately and lack context, more recent developments, such as those reported by Arora et al.\cite{Arora21}, present imitation methods that use anchored projection techniques. However, the mimicry approach uses a 3D mesh builder that attaches volumetric blocks to our model's surface, leading us to opt for Texture Painting\cite{Shahrabi21} instead. While mimicry forms a foundation for our drawing techniques, it struggles with engineering applications that require frequent model scaling. These limitations prompt further consideration and refinement of our techniques.

\subsection{Integrating bitmap drawing and vectorization for Enhanced Precision  in 3D AR Environments}
In 2018, Dudley et al.\cite{Bare-Handed-3D-Drawing} proposed a method for drawing vector lines in 3D mid-air, using a technique known as 'Tapline', which involves plotting points in space but is quite time-consuming. 
Our solution requires a method for directly vectorize curves on 3D surfaces.Existing 2D vectorization methods developed by Noris et al.\cite{Noris13} and Donati et al.\cite{Donati17, Donati19} often use 1-skeleton or raw image gradients to highlight important drawing features such as curve endpoints and junctions\cite{Najgebauer19,Bo15,Donati17,Donati19,Noris13}. Although these methods are effective in clean line drawings, they can suffer from geometric and topological errors under noisy conditions. The introduction of the PolyVector Flow method by Puhachov et al.\cite{Vectorization-Line-Drawing} marks a significant improvement by efficiently converting raster images to vector formats, focusing on tasks such as identifying crucial points, extracting topology, and determining optimal geometries. However, it is time-consuming and primarily suitable for line art. The VTracer method\cite{Sanford20} integrates these processes by initially clustering the input image using Hierarchical Clustering\cite{Chris20}, then tracing each cluster into a vector. The vector tracing algorithm involves converting pixels into paths, simplifying these paths into polygons, and smoothing the polygons with a curve-fitting technique.This approach is particularly adept at handling complex junctions in noisy environments and underpins our strategy to enhance 3D vectorization. By extending these principles to 3D, our project aims to improve the accuracy and quality of drawing in AR. After the vectorization process, the line records stroke data as mathematical vectors, enabling tasks such as curve adjustments and maintaining clarity and quality when scaled. This makes it suitable for use in other engineering software.

\subsection{Semi-Structured Interviews as Qualitative Metrics}
Semi-structured interviews play a critical role in qualitative research, enriching the data gathered with insights and depth not typically accessible through quantitative methods alone. These interviews typically employ a two-tiered approach to questioning:  main themes and further follow-up questions. The main themes serve as the framework for the conversation, addressing the major questions of the research and enabling participants to openly express their opinions and experiences.It is customary to engage every participant in these primary themes\cite{Astedt-Kurki94}. The sequence of these themes is often designed to progress in a logical and orderly manner\cite{krauss09}, starting with more general questions to ease participants into the conversation and establish a comfortable atmosphere\cite{Whiting08,Cridland15}. The initial questions typically touch on areas familiar to the participants but crucial to the research topic\cite{Whiting08}. As the interview progresses, the discussion deepens, moving towards more comprehensive topics\cite{Whiting08,baumbusch2010,Cridland15}  before returning to less intense themes towards the end\cite{baumbusch2010}. By incorporating semi-structured interviews into our study, we are hoping to gather an extensive variety of information, from user interactions with technology to the subjective experiences that demonstrate its applicability and user approval.

\section{System Design}
To meet the demands of drawing on 3D geographical models, we have developed hand-interaction-based drawing techniques utilizing the Microsoft HoloLens2\cite{hololens}. This augmented reality headset, equipped with depth cameras and sensors, allows for precise environmental sensing. Our system enables users to draw freely by projecting their fingertips' positions onto 3D surfaces using a ray. The resultant drawings are then vectorized by tracking the points in the stroke data to form vectorized lines, which are then smoothed. This integration is powered by the Unity game engine\cite{unity}, which facilitates a smooth interaction between drawing functionalities and 3D modeling. The system’s innovation lies in its capability to vectorize 3D stroke data. The bitmap drawing method enables users to swiftly sketch designs or annotate directly on models, and the vectorized stroke is scalable and editable, ensuring no loss of detail upon scaling or exporting.

\subsection{Workflow Description}
The workflow is optimized for user efficiency and accuracy, encompassing several key phases to ensure a smooth operation from start to finish:

\begin{figure}
    \centering
    \includegraphics[width=1\linewidth]{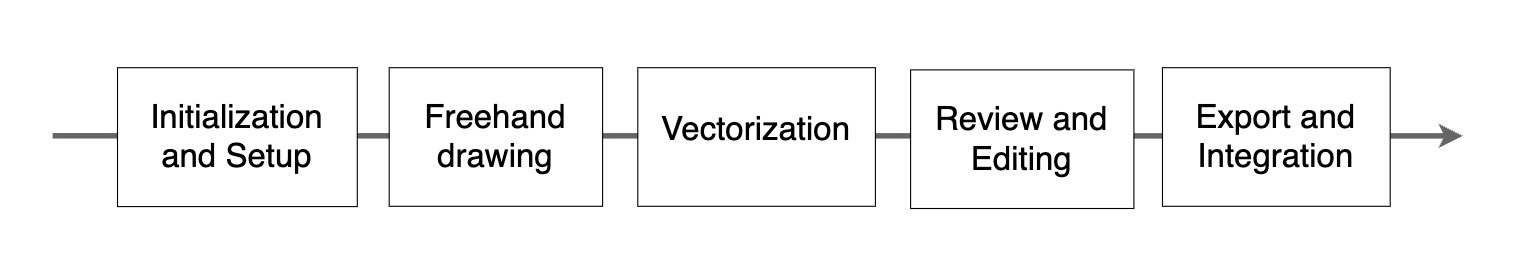}
    \caption{Draw2Vector Workflow}
    \label{fig:Workflow}
\end{figure}

\begin{itemize}
\item \textbf{Initialization and Setup:} 
To begin, users load a 3D model from the application's cloud-based library. The HoloLens2 headset runs system checks and measurements to ensure proper tracking and rendering based on the user's location.

\item \textbf{FreeHand Drawing:} Users perform drawing tasks on the surface of 3D scanned models using gesture controls, providing an immersive experience. The system supports various function like changing the size and color of the brush, accessible via a UI panel in the AR view.

\item \textbf{Vectorization:}
Users can convert their freehand sketches into vectorized strokes by clicking the 'Vectorization' button on the UI panel. This process optimizes the paths of the sketches, enhancing clarity and reducing noise to produce high-quality 3D vectorized stroke. 

\item \textbf{Review and Editing:} 
If the vectorized strokes are not satisfactory, they can readily edit the vectorized drawings. This editing phase includes editing the key points of the strokes, allowing users to achieve the goal they desire for the final output.

\item \textbf{Export and Integration:} Completed drawings are exported as a CSV file that records the geographic coordinates of each point and includes a texture map of the model. These files are suitable for use in other applications for further editing and refinement.

\end{itemize}

Each of these steps is designed to ensure that the system is not only user-friendly but also capable of producing high-quality outputs that are useful in professional settings. Next, we will introduce the two most important steps in our workflow: Free-hand drawing and the Vectorization process. This workflow \ref{fig:Workflow} provides a visual summary of the entire process.

\subsection{User interface} 
\begin{figure}
    \centering
    \includegraphics[width=0.5\linewidth]{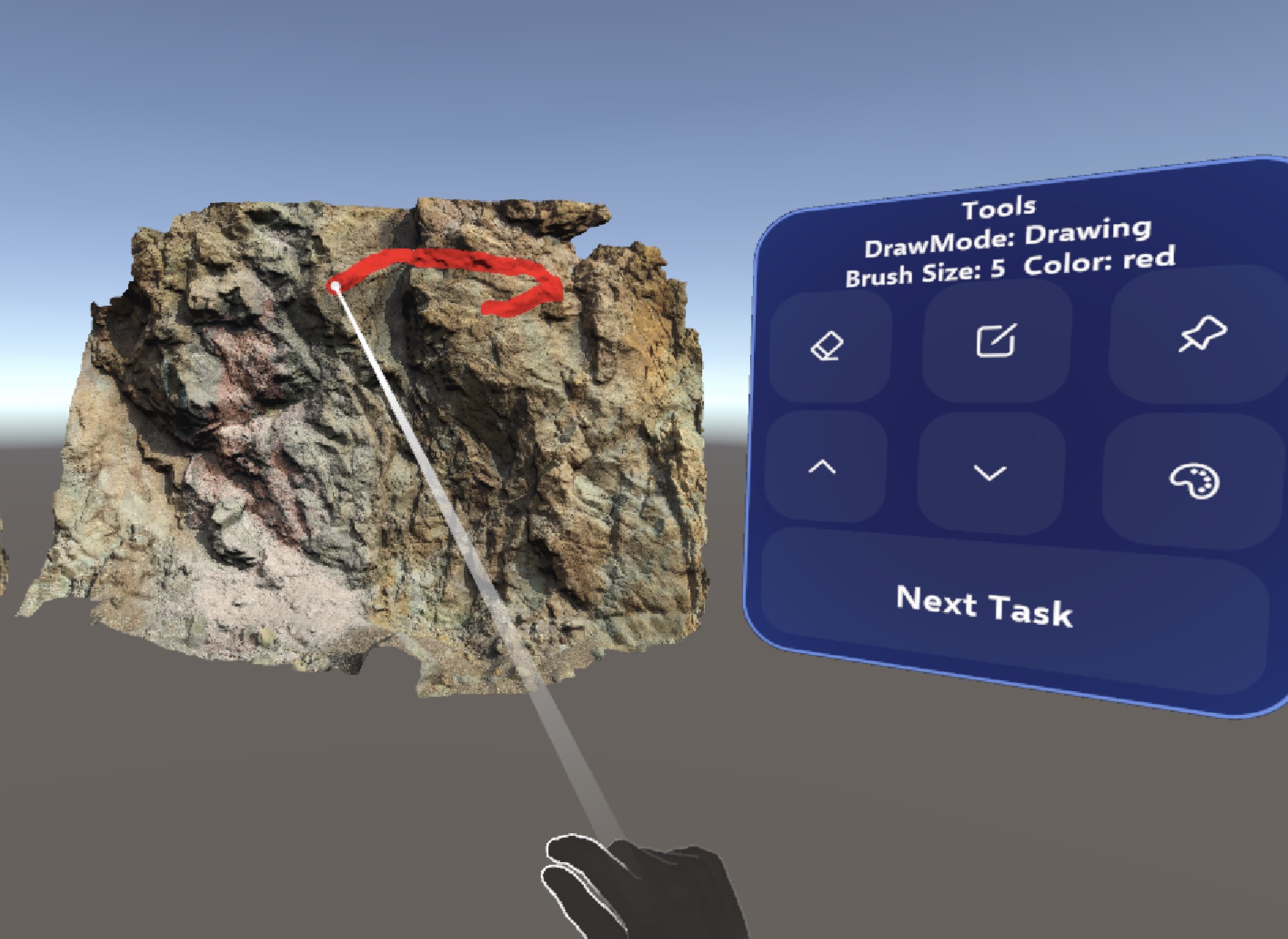}
    \caption{User interface of Draw2Vector}
    \label{fig:UI}
\end{figure}

Upon loading the model, the system automatically enters the freehand (bitmap) drawing mode. A small red sphere, representing the user's hand position, appears on the surface of the 3D model. Drawing begins with a pinch gesture in the air. As the user performs this gesture and move his hand, the red sphere moves along the model’s surface, tracing the strokes. The drawing process concludes when the gesture is released. Users can navigate the red sphere to the right-side UI panel\ref{fig:UI} and pinch again to activate the panel, which appears directly in front of the user. This panel provides several options, including adjusting brush color and size, clearing the canvas, vectorization, and changing the drawing mode.

If the user selects the vectorization option, the current freehand drawing is converted into vectorized lines. Subsequently, the sphere changes to green, indicating the Tapline mode (vector drawing mode). If users are unsatisfied with the lines, they can adjust the position of points on the line or continue to draw using vector techniques by placing points on the figure that automatically connect into lines.

It's straightforward for users to stay updated with their current settings because the top of the UI banner displays the drawing mode and brush size.The user also has the option to switch from freehand drawing to Tapline mode if they would like to modify the drawing mode. Their work is automatically saved and exported after the drawing session is finished.

\begin{figure}
    \centering
    \includegraphics[width=0.3\linewidth]{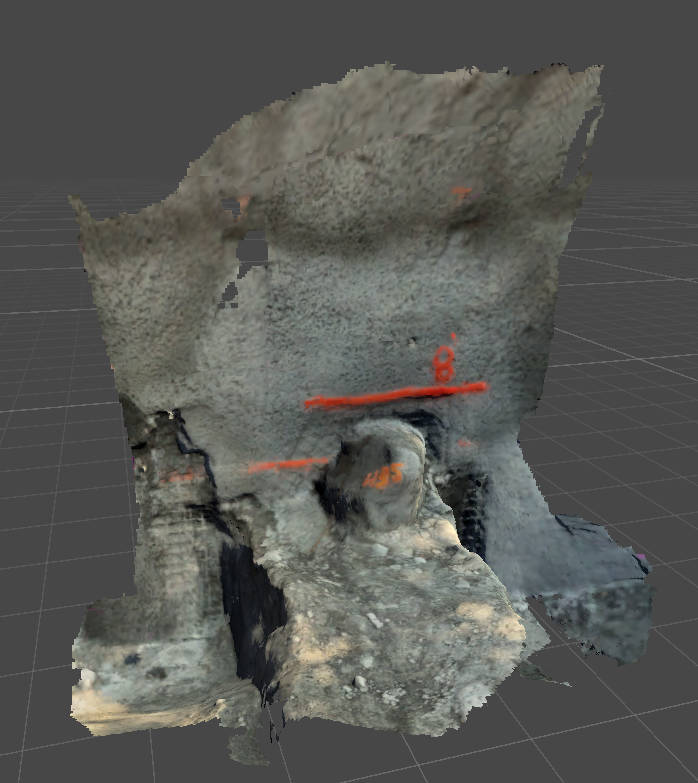}
    \caption{Making annotation}
    \label{fig:making annotation}
\end{figure}

\subsection{Freehand Drawing: Quick sketch on the surface of 3D scanned models} 
Freehand Drawing is specifically designed for rapid sketching directly on 3D models. This function is extremely useful in a collaborative scenario like a virtual meeting. In such scenario, the host can show to other participants easily where they are going to modify. This tool can also be used as an annotation tool, which allows engineers to point out where needs to be, as shown in this photo\ref{fig:making annotation}, excavated.

\begin{figure}
    \centering
    \includegraphics[width=1\linewidth]{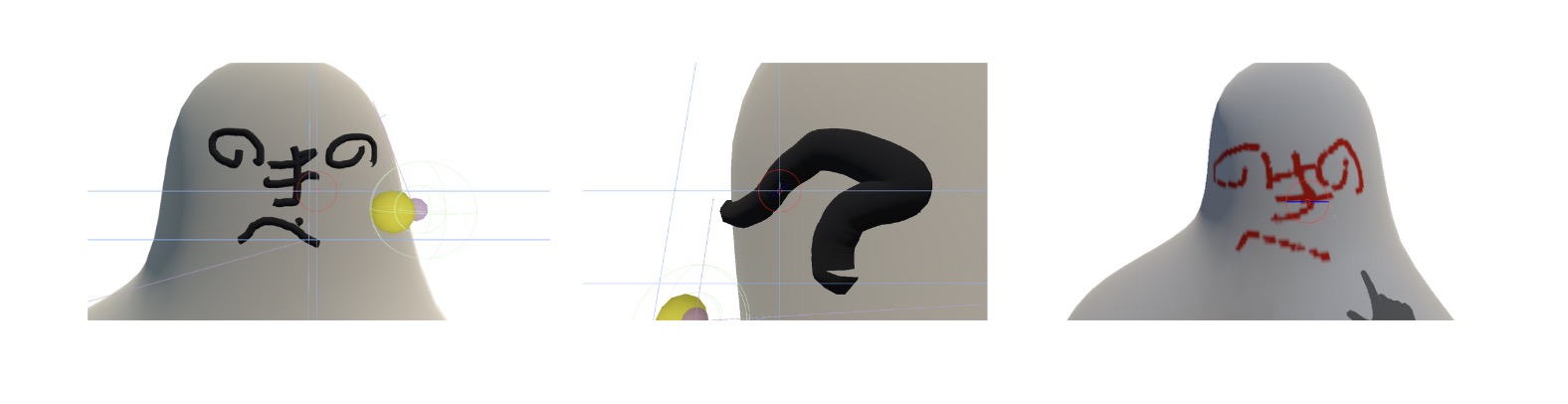}
    \caption{Comparison between Mimicry - 3D mesh builder(left) and Texture Painting (right)}
    \label{fig:comparison}
\end{figure}

\textbf{Mimicry method:}
In this research project, we will use a context-based drawing approach called \textit{mimicry}\cite{Arora21}, this method is called \textit{mimicry} because the user tends to \textbf{mimic} the shape of their intended curve on surface. Such drawing algorithms are used to project 3D curves, in our research, the stroke on 3D surface. Traditional projection algorithms \textit{Spraycan} method tend to have the limitation that can't capture the model curve, especially when models have complicated, layered surfaces. So here we will introduce the technical description of this method. The core principle of the mimicry method is to accurately map the trajectory and curvature of freehand strokes made in mid-air onto a virtual object's surface. The mathematical framework involves the projection of stroke points onto the 3D model and optimization of the resulting curve for smoothness and continuity.

\begin{itemize}
    \item \textbf{Projection of Stroke Points:} For subsequent points \( p_i \) (where \( i > 0 \)) of the stroke in mid-air, each point is projected onto a corresponding point \( q_i \) on the surface \( S \) of the 3D model. In our research, we utilize both the previous stroke data \( p_{i-1} \) and the previous projection \( q_{i-1} \) to enhance the projection accuracy. This process is mathematically defined by the following equations:

    \[
    r_i = q_{i - 1} + \Delta{p_i},
    \]

    where 
    \[
    \Delta{p_i} = p_i - p_{i-1}
    \]

    \[
    q_i = \arg \min_{x \in S} \|r_i - x\|
    \]
    where \( \|r_i - x\| \) is the Euclidean distance between \( r_i \) and \( x \).
\end{itemize}

\begin{figure}
    \centering
    \includegraphics[width=0.3\linewidth]{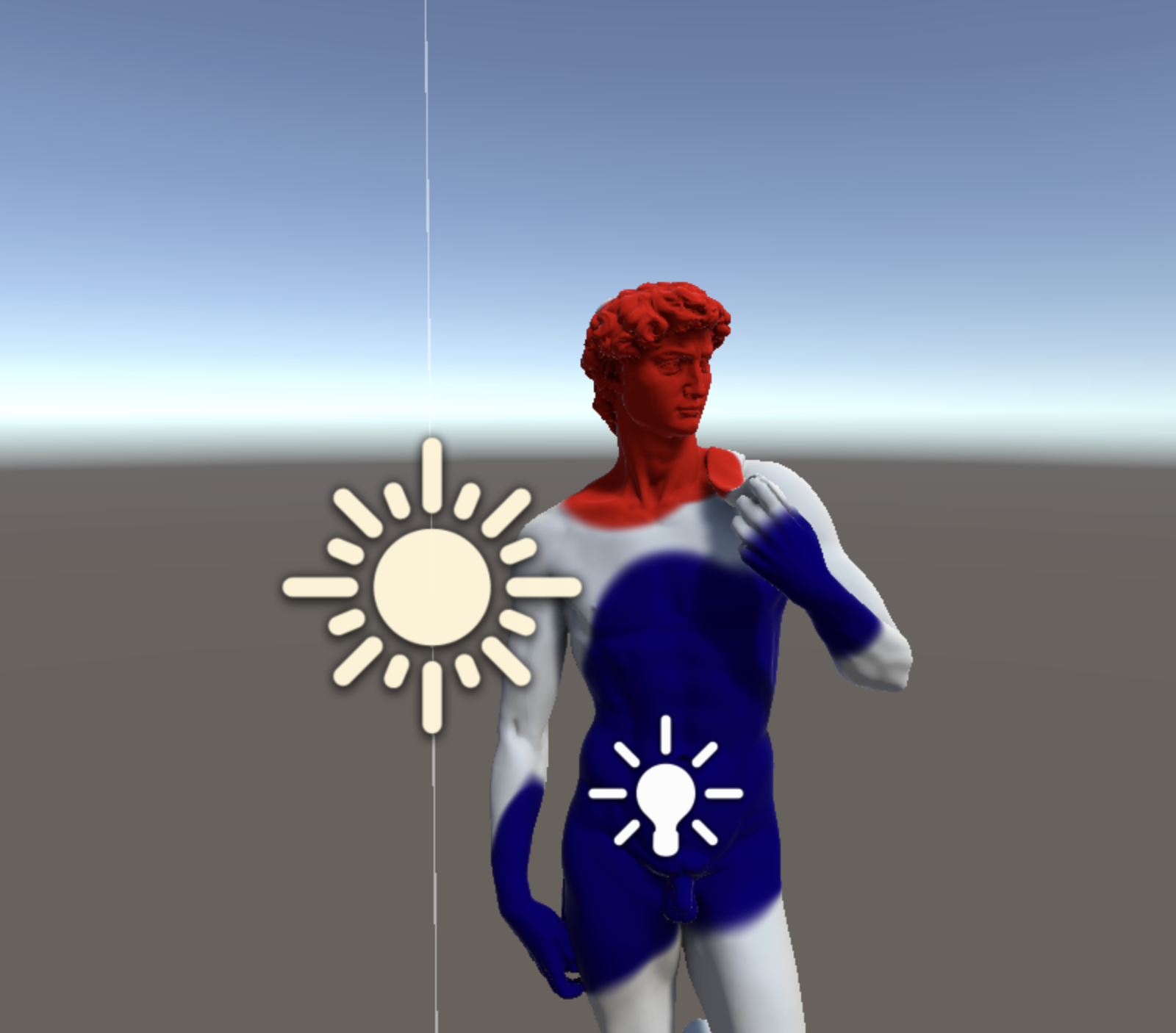}
    \caption{Texture Painting method}
    \label{fig:texture painting}
    \cite{Shahrabi21}
\end{figure}

\textbf{Texture Painting Method:}
During the implement our project, we encountered specific challenges with the initial implementation of the mimicry method for freehand drawing on 3D models. Originally, the mimicry approach relied on a 3D mesh builder for drawing, which attaches a volumetric block to the surface\ref{fig:comparison}(Figure 4), is unsuitable for our project's needs. To address this, we have shifted to using Texture Painting\ref{fig:texture painting}\cite{Shahrabi21}(Figure 5). 

Texture Painting is a technique we have implemented to enhance the drawing capabilities within our project. This method involves creating separate 2D image files for each layer, which are then bound to the object's existing texture and material (Mtl) files. This integration allows for the textures to be updated in real-time and adhered directly to the 3D surfaces, significantly enhancing both the flexibility and performance of the drawing tool. Texture Painting leverages shaders to dynamically apply colors or textures to a mesh based on real-time user interactions\cite{Shahrabi21}, such as clicking or dragging over the mesh surface. The shader used in this process is designed to project a texture onto a mesh while modifying the texture's pixels directly on the mesh. This allows for immediate visual feedback and artistic control, enabling users to see the effects of their modifications instantly. This capability is crucial for applications requiring high levels of interaction, as in our project where responsive freehand drawing on 3D models is essential.

\textbf{Handling Discontinuous Lines During Rapid Drawing: }
To address the issue of discontinuous lines when drawing quickly, we implemented an improvement in our freehand drawing method. If the distance between points from consecutive frames is greater than the brush size, the line will appear discontinuous. To solve this, if the distance between the points is greater than the brush size, we connect these points by projecting the line segment onto the 3D object's surface texture, ensuring continuous lines. This enhancement ensures that rapid strokes are accurately represented as continuous lines, maintaining the integrity of the drawing experience.

\textbf{Summary: }
We initially implemented the Projection of Stroke Points technique from the mimicry method, which utilizes a 3D mesh builder. This approach effectively attaches a volumetric block to the surface, making it less suitable for engineering tasks such as drawing, annotating, and coloring. Therefore, we transitioned to using texture painting for rendering our drawings. This adaptation allows for a more appropriate application of markings and colors directly onto the 3D model's surface, enhancing the usability of our tool in engineering contexts.

\subsection{Vectorization Method Evolution} 
Initially, the project used the \textit{VTracer} method \cite{Sanford20} to transform hand-drawn movements into vector strokes. Originally designed for 2D graphics, \textit{VTracer} rapidly converts strokes to vectorized data, ensuring scalability without loss of information. 

The vector tracing process begins by applying Hierarchical Clustering\cite{Chris20} to the input image, which segments it into distinct clusters. Each cluster is then converted into vector format through a three-step algorithm: First, the pixels within each cluster are transformed into a continuous path. Next, this path is simplified into a polygon by minimizing the number of vertices to maintain the shape's integrity. Finally, the polygon is smoothed and refined using a curve-fitting technique to enhance the visual quality and accuracy of the resulting vector image.

We implemented this vectorization technique specifically targeting the texture images of 3D models. The process begins by mapping sketches directly onto the separate texture layers of these models. Once mapped, these texture images go through a detailed vectorization process. This involves converting the 2D representations of the textures into vector graphics. The converted vector graphics are then re-applied to the 3D models and subsequently evaluated within an augmented reality (AR) environment. This two-phase approach—first vectorizing the textures in 2D and then integrating them back onto the 3D models—ensures that the vectors retain true to the model's detailed structure when viewed in AR. This method allows us to maintain precision in how textures map onto the complex contours and features of the 3D models, ensuring that the final visualizations in AR are both accurate and visually coherent.

However, after implementing this method, we identified significant limitations. 

\begin{itemize} 
\item Primarily, the technique outlines images rather than capturing them as continuous, dynamic vector lines. Consequently, while the graphics accurately trace the external boundaries of shapes, they do not reflect the true nature of hand-drawn strokes. 

\item Moreover, the current vectorization technique does not support modifications in an AR environment, restricting its interactive potential, which is crucial for real-time applications and field adjustments.
\end{itemize} 

\textbf{Refinement of the 3D Vectorization Process:}
The initial shortcomings of the \textit{VTracer} method led to a comprehensive revision of our vectorization approach. To better preserve the intrinsic qualities of hand-drawn strokes, we developed an advanced vectorization technique that incorporated real-time tracking of the drawing's 3D coordinates. This enhancement allowed for a more nuanced capture of each stroke's dynamic properties, ensuring that the vectorized outputs retained the fluidity and expressiveness of the original sketches.

\begin{figure}
    \centering
    \includegraphics[width=0.8\linewidth]{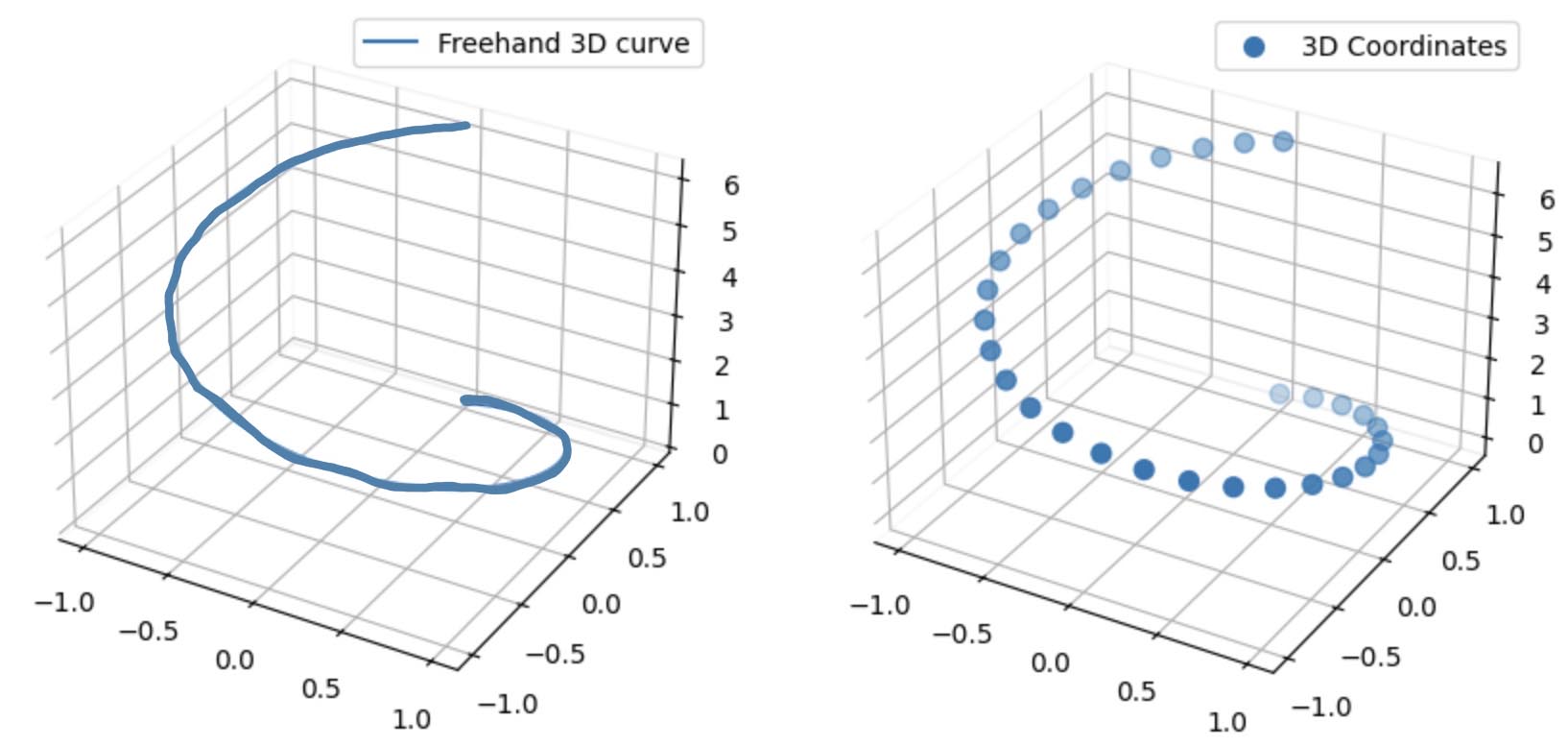}
    \caption{Tracking 3D Coordinates: Left - Continuous Freehand Drawing; Right - 3D Points Capturing the Stroke Path}
    \label{fig:tracking 3D Coordinates}
\end{figure}

\begin{enumerate}
\item \textbf{Tracking 3D Coordinates: }
During live drawing sessions, we begin our enhanced vectorization process by properly recording the 3D coordinates of points as they are projected onto a curve. This tracking occurs in real time and captures every detail of the user's drawing motion. By recording these locations at regular times, often every few milliseconds, we assure a detailed description of the drawing trajectory, which is used to generate exact 3D vectors\ref{fig:tracking 3D Coordinates}(Figure 6).

\begin{figure}
    \centering
    \includegraphics[width=0.4\linewidth]{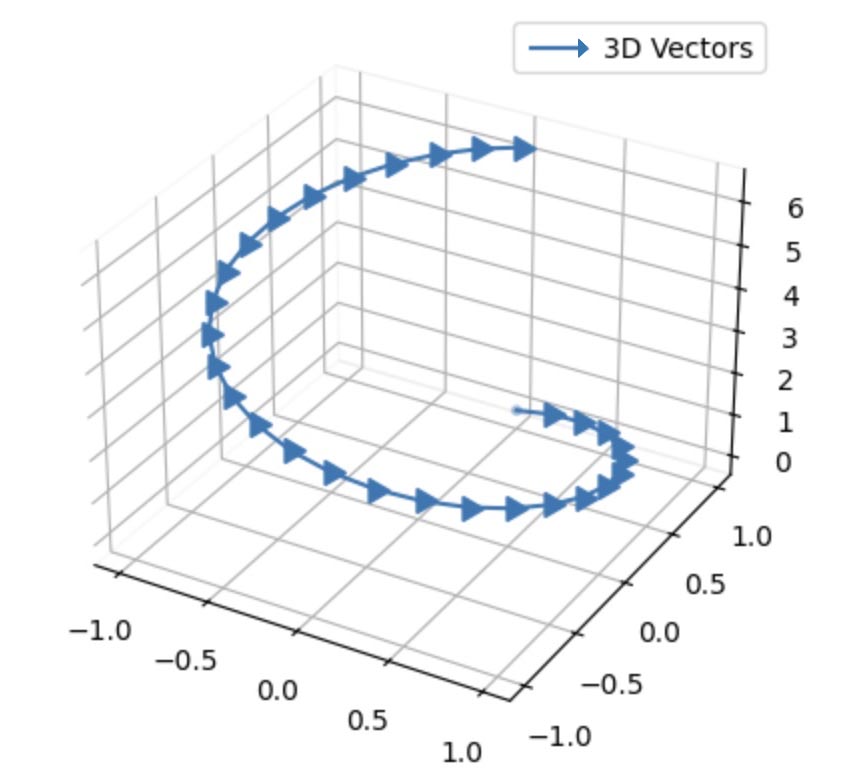}
    \caption{Joining Points to Form Discrete Vectors}
    \label{fig:joining Points}
\end{figure}

\item \textbf{Joining Points to Form Discrete Vectors: }
Once the points along the curve have been traced, the next step is to combine them to form discrete three-dimensional vectors\ref{fig:joining Points}(Figure 7). This procedure converts the sequence of points into a collection of vectors that follow the route of the original drawing in 3D space. Although this results in a segmented line, it is an important initial step in developing a continuous vector representation.

\begin{figure}
    \centering
    \includegraphics[width=0.8\linewidth]{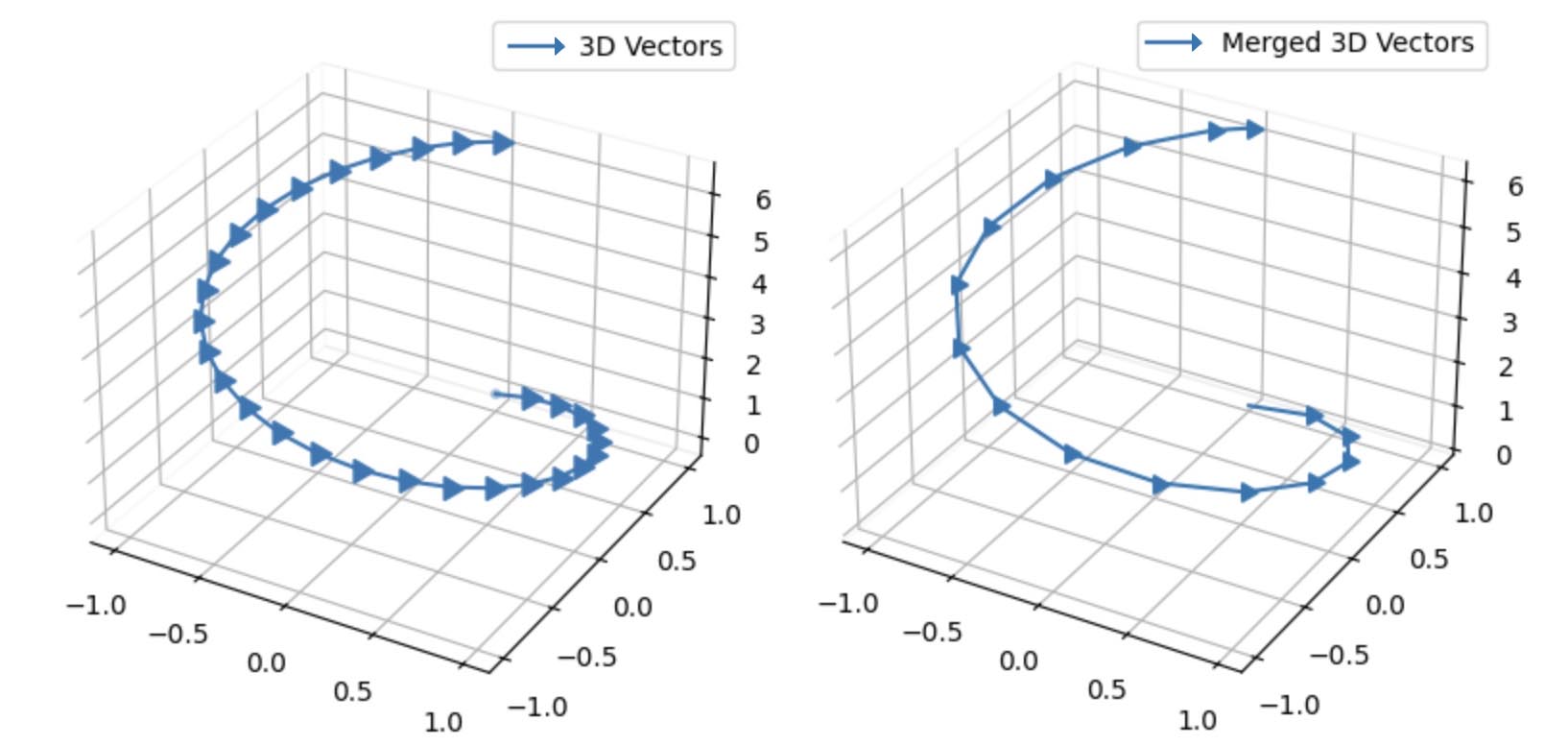}
    \caption{Path Walking: Left - 3D Vectors; Right - Merged 3D Vectors}
    \label{fig:path Walking}
\end{figure}

\item \textbf{Path Walking: }
To improve the continuity between these discrete vectors, we use a technique called Path Walking\ref{fig:path Walking}(Figure 8). This method focuses on merging vectors that are all in the same direction. By combining these aligned vectors, Path Walking eliminates abrupt directional changes and contributes to a more unified and flowing vector path. This approach is essential for preserving the original drawing's integrity and smoothness in vectorized form.

\begin{figure}
        \centering
        \includegraphics[width=0.8\linewidth]{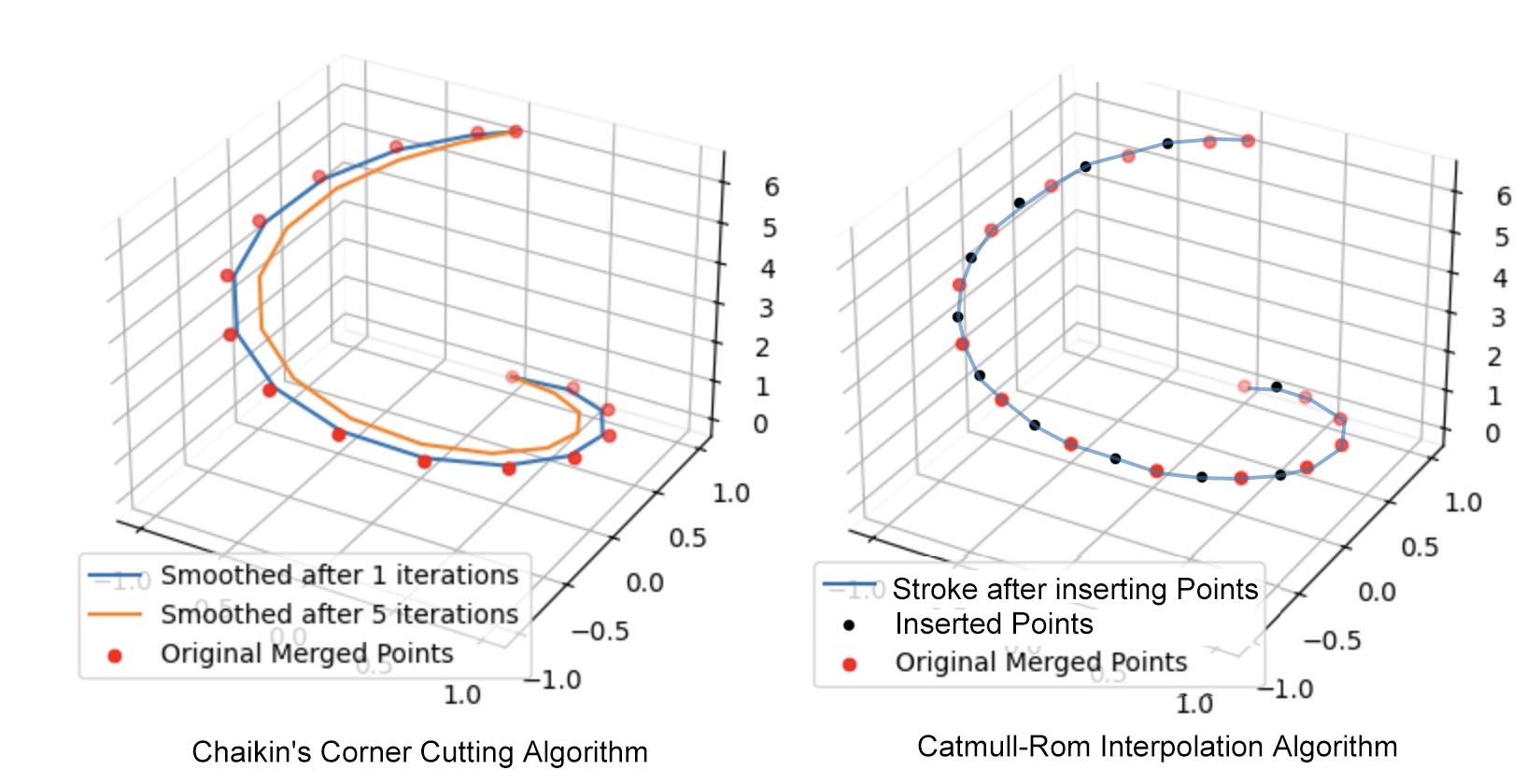}
        \caption{Left: Significant deviation resulting from Chaikin's Corner Cutting Right: Shift to using Catmull-Rom Interpolation Algorithm}
        \label{fig:chaikin-deviation}
    \end{figure}

\item \textbf{Path Smoothing: }
The final phase in our method is path smoothing, depicted in Figure 9\ref{fig:chaikin-deviation}, which refines the vector path to produce a smoother, more continuous line. Initially, we employed Chaikin's Corner Cutting Algorithm\cite{CHAIKIN1974346}. However, we observed that repeated iterations of this algorithm resulted in significant deviations. Consequently, we shifted to using Catmull-Rom Interpolation Algorithm\cite{CATMULL1974317}. This method ensures that the curve smoothly passes through the control points, thus maintaining the fidelity of the path. It is described by the following equation:
\[
P(t) = 0.5 \left( 
(2 p_1) + 
(-p_0 + p_2) t + 
(2 p_0 - 5 p_1 + 4 p_2 - p_3) t^2 + 
(-p_0 + 3 p_1 - 3 p_2 + p_3) t^3 
\right)
\]

The parameter \( t \) represents the interpolation factor between points, ranging from 0 to 1, with \( t=0 \) at \( p_1 \) and \( t=1 \) at \( p_2 \). This parameterization allows for precise control over the interpolation process, enabling adjustments for greater accuracy or visual appeal in the resultant vector path. In our application, given the scale of our drawings, we currently default to inserting only one point between each pair of points. However, this parameter can be adjusted to achieve greater precision if needed.

\item \textbf{Path Projection: }
When constructing a continuous line on the surface between two points, we can utilize the following procedure to ensure accuracy and continuity:

\begin{itemize} 
\item Plane Construction: Given two points and the Raze direction (the ray pointing from the finger to the model), we construct a plane. The raze direction, along with the two points, helps define this plane.
\item Intersection Calculation: Using this plane, we calculate the intersection line. The raze direction and the two points form a triangle. The intersection of this triangle with the 3D model surface determines the path we need to project. By identifying the intersecting segment of this triangle with the model's surface, we can accurately color this segment, ensuring a continuous and seamless line on the 3D object.
\end{itemize} 

\end{enumerate}

\subsection{Implement details}
Our development is based on the official MRTK(Mixed Reality toolkit) template\cite{Microsoft20}, and our techniques are implemented in C\# and Shader, utilizing the Unity Engine for interaction, rendering, and VR support, including integration with Hololens2. During the freehand drawing phase, we combine the mimicry method\cite{Arora21} and texture painting techniques\cite{Shahrabi21}. In the vectorization stage, we employ the key idea of the VTracer method\cite{Sanford20}, such as path walking, and we utilize Catmull-Rom Interpolation Algorithm\cite{CATMULL1974317} to achieve path smoothing. This comprehensive approach enhances the usability of 3D drawing, significantly improving collaboration and design processes in mixed reality environments.

\section{User Study}
We conducted a user study to evaluate and compare the performance of data from the freehand drawing phase and post-vectorization, as well as user experience with the system. The study involved developers and engineers, aiming to enable participants to perform drawing tasks on predefined 3D models, observe their workflows, and store the results.

\subsection{Participants}
We recruited 10 participants (5 female, 5 male) with a mean age of 38.75 (SD = 13.34). The participants were a mix of engineers, who are the actual users of the system, and software developers, who provided valuable feedback on the system's usability and functionality. Six of these participants reported prior experience with the HoloLens 2, which was used for the experiments.

\subsection{Procedures}
Participants began the experiment with a briefing that introduced them to the AR headset and explained the drawing techniques. They then completed a familiarization task to ensure comfort and proficiency with the system. Subsequently, they engaged in a series of drawing tasks designed to evaluate the practical application and effectiveness of the techniques. After the freehand drawing phase, the drawing data were stored. After that, participants used a vectorization function to convert bitmap strokes into vectors. If participants were unsatisfied with the vectorized lines, they could then adjust the points of the lines. Data was collected again. Participants provided feedback through semi-structured interviews and questionnaires, assessing user satisfaction, ease of use, and system responsiveness, among other factors. Following the experiment, we compared bitmap and vectorized lines in terms of line smoothness, accuracy, data size and so on. This method enabled a full evaluation of both technical performance and user experience, which enabled an extensive examination of the system's effectiveness.

\subsection{Training and familiarisation task}
Participants were given a thorough training that introduced the AR headset and shown how to use the platform and its drawing skills. Before starting the main tasks, participants completed a gesture familiarization exercise involving a rock model dotted with ten points. Participants were instructed to connect these dots in a continuous line, practicing multiple times until they could complete the task within 20 seconds.

\begin{figure}
    \centering
    \includegraphics[width=0.8\linewidth]{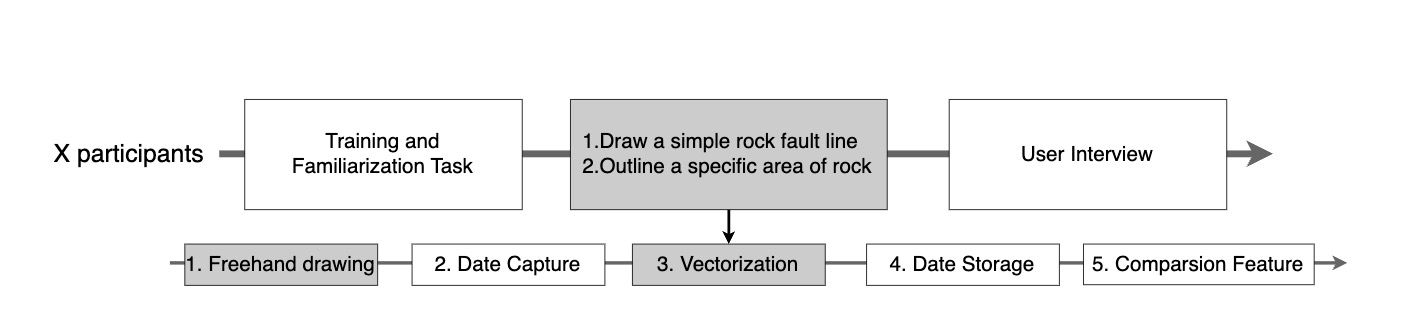}
    \caption{The illustration of the protocol}
    \label{fig:enter-label}
\end{figure}

\subsection{TASK1: Draw a Simple Rock Fault Line}

\begin{figure}
    \centering
    \includegraphics[width=0.5\linewidth]{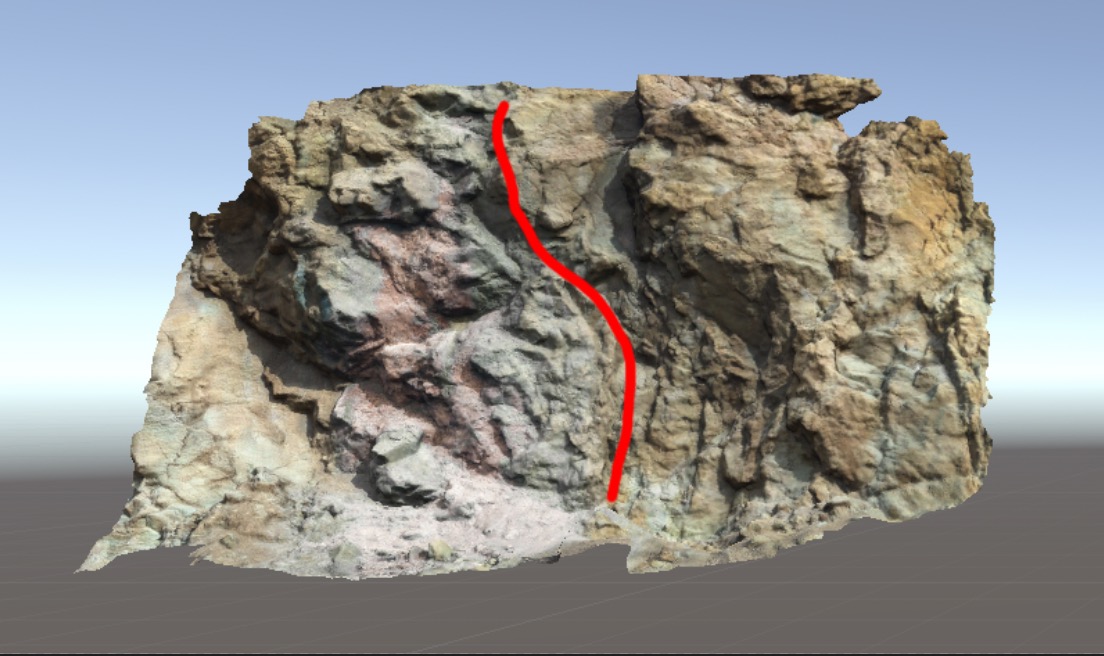}
    \caption{Draw a Simple Rock Fault Line}
    \label{fig:Rapid line Drawing}
\end{figure}

Participants will be presented with a 3D rock model displaying a prominently marked geological fault line\ref{fig:Rapid line Drawing}, originally delineated by engineers in a digital model. This thick line represents a fault and serves as a visual guide to clarify the task's objective. The surface of the rock model designated for drawing is relatively flat, and the entire fault line is visible within the user's field of view, making the task straightforward. Participants are tasked with replicating this fault line as efficiently as possible on an identical rock model positioned adjacent to the original, without the need for adjusting their viewpoint significantly.They can make necessary adjustments during the drawing process, the task is considered complete once the participant is satisfied with their replication. This task allows participants to quickly and easily complete the drawing, reflecting the simplicity and directness of the task.

\begin{figure}
    \centering
    \includegraphics[width=0.5\linewidth]{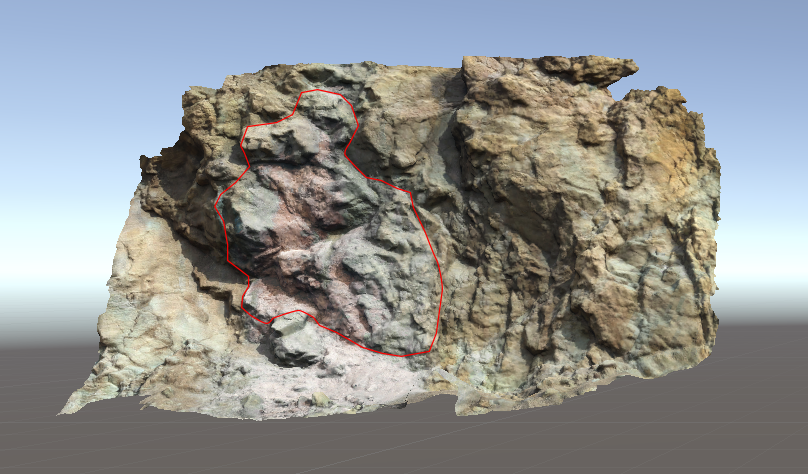}
    \caption{Outline a Specific Area of Rock}
    \label{fig: Enclosed Precise Feature Drawing}
\end{figure}

\subsection{TASK2: Outline a Specific Area of Rock}
In this task, participants will observe a rock model where a specific section of a geological layer is outlined with a clear closed line\ref{fig: Enclosed Precise Feature Drawing}, originally marked by engineers in a computer model. Their objective is to replicate this enclosed line on a similar rock model located next to the original. The outlined area is relatively large, requiring participants to adjust their viewpoint or adjust their viewpoint to complete the drawing accurately. This task tests the participants' ability to draw and delineate an enclosed area on a 3D rock surface, emphasizing accuracy and the ability to manipulate the model for better visibility.

\subsection{Vectorization of Drawn Lines}
Upon completing each line drawing, users can vectorize the lines through the UI panel. Selecting this option initiates the system's automatic conversion of the originally sketched line into a vectorized curve. This vectorized curve is smooth, aesthetically pleasing, and editable.In vectorized mode, users can select and modify points along the line to alter its shape, achieving their desired outcome.

\subsection{Questionnaire and the Semi-structured User Interview}

After completing the tasks, participants fill out a questionnaire. This questionnaire collects immediate feedback on their experience, focusing perceived responsiveness, editable satisfaction, and scalability satisfaction. Following the completion of the questionnaire, participants undergo a semi-structured interview\cite{baumbusch2010}. This interview allows for a more in-depth exploration of their experiences, opinions, and suggestions for improvement, building on the responses given in the questionnaires. These interviews provided insightful input for upcoming improvements by attempting to comprehend the complex viewpoints and prospective enhancements from the user's point of view.The interview was organized using a series of pre-planned open-ended questions that allowed participants to share their opinions and experiences while also serving as a guide to ensure that all relevant subjects were covered. The interviews explored several key areas:

\begin{itemize}
\item \textbf{Perceived Responsiveness}: "How would you rate the responsiveness of the AR system during the drawing tasks? Did the system react quickly to your inputs? "
\item \textbf{Editable Satisfaction}: "After vectorizing the lines, how satisfied were you with the ability to edit the vectorized curves? Did the editing features meet your expectations for flexibility and precision?"
\item \textbf{Scalability Satisfaction}: "After scaling the drawings, how would you assess the clarity and detail retention? Were the details preserved to your satisfaction?"
\item \textbf{Functionality and Tools}: "Which functions or tools within the system did you find most useful or lacking? How did these impact your drawing process?"
\item \textbf{System Usability}: "How did you find the usability of the AR system for drawing tasks? Were there any aspects that felt particularly intuitive or challenging?"
\item \textbf{Overall Experience}: "Can you describe your overall experience using the AR drawing system? What were the highlights and lowlights?"
\item \textbf{Suggestions for Improvement}: "What improvements would you suggest for enhancing the drawing experience in this AR system?"
\item \textbf{Future Usage}: "Based on your experience, how likely are you to use this type of technology in your regular work? What changes would make it more appealing or useful for you?"
\end{itemize}

With the participant's permission, audio recordings of the around 10-15 minute interviews were made in order to ensure accuracy in the gathering and processing of data. The interviewer followed the guide but also took the opportunity to go deeper based on the participant's responses, ensuring a comprehensive understanding of their experience.

\section{Data Analysis}
The data analysis phase is crucial for interpreting the results gathered from the user study and drawing meaningful conclusions. This chapter outlines the procedures and techniques used to analyze the data collected from the questionnaires, tasks, and semi-structured interviews. The analysis is divided into quantitative and qualitative sections to provide a comprehensive understanding of user experiences and system performance.

\subsection{Quantitative Analysis}
Quantitative data were collected from questionnaires and task performance metrics. The following quantitative metrics were analyzed:

\textbf{Quantitative Metrics: }These metrics are directly measured from the data collected during the tasks and questionnaires:

\begin{itemize}

\item \textbf{Impact of Vectorization on Data Size}: Evaluates and compares the file sizes of the lines from freehand stage and vectorization stage. This metric helps understand the impact of vectorization on data efficiency and storage requirements.

\item \textbf{Impact of Vectorization on Curve Smoothness}: This metric assessed the fluidity and visual quality of lines from the freehand stage and vectorization stage, indicating how natural and smooth the lines appeared on the 3D models. This metric can be quantitatively assessed using angle difference:

\textbf{Angle Difference}: To quantitatively analyze the smoothness, the angle changes between adjacent segments on two stages of the curve are examined. The coordinate of each vertex is recorded in a CSV file before and after vectorization. This involves computing the angle between tangent vectors at successive points, which can be determined using the dot product formula:

\[
\cos(\theta) = \frac{T_1 \cdot T_2}{|T_1| |T_2|}
\]
where \( T_1 \) and \( T_2 \) are the tangent vectors at consecutive points along the curve, \( \cdot \) denotes the dot product, and \( |T_1| \) and \( |T_2| \) are the magnitudes of these vectors. This calculation provides the cosine of the angle \( \theta \), indicating the directional change of the curve at each point. 

\item \textbf{Impact of Vectorization on Curve Accuracy}: This metric evaluates the similarity between lines from the freehand stage and the vectorization stage by measuring the deviations of the vectorized lines from the freehand lines. This analysis helps identify which stage retains more detail and precision in drawings on 3D models. To measure accuracy, the vectorized curves can be superimposed on the freehand curves, and deviations can be quantified by calculating the Euclidean distance between corresponding points on the freehand and vectorized curves.

\item \textbf{Perceived Responsiveness:} During drawing tasks, participants rate the perceived responsiveness of the AR system on a scale of 1 (very unsatisfactory) to 5 (extremely satisfactory). This input facilitates the assessment of the system's functionality in real-time applications.

\item \textbf{Scalability and Detail Retention:} Following the scaling of drawings, participants rate the effectiveness on a scale of 1 (very poor) to 5 (excellent), depending on how well details are retained. This measure assesses the system's adaptability to changes in scale in terms of quality and detail.

\item \textbf{Editable Satisfaction}: Participants rated their satisfaction with the ability to edit vectorized lines on a scale of 1 (very unsatisfied) to 5 (very satisfied). This measure assessed how well the editing features met their expectations for flexibility and precision.

\end{itemize}

\subsection{Qualitative Analysis}
Qualitative data were collected from semi-structured interviews conducted after the tasks. These interviews provided deeper insights into user perceptions and experiences. The following qualitative metrics were analyzed:

\textbf{Qualitative Metrics:}
These metrics are derived from participants' subjective evaluations, providing insights based on user perception and experiences:

\begin{itemize}
\item \textbf{Total User Experience:} Participants highlight both the greatest and least rewarding parts of the system as they discuss their overall experience with it. This measure pinpoints particular high and low times in the system's use and assesses user satisfaction.

\item \textbf{System Usability and Functionality:} Users evaluate the AR system's general usability for drawing tasks and talk about the usefulness of certain tools and functions. This includes comments on any easy-to-understand or difficult-to-understand elements they come across that affect their workflow and drawing process.

\item \textbf{Suggestions for System Improvement:} Participants offer direct feedback on how to improve the drawing experience in the augmented reality environment, along with possible system changes.

\item \textbf{Future Usage and Appeal:} Participants consider how likely they are to utilize this technology on a daily basis at work and offer suggestions for improvements that could make the system more user-friendly and appealing. This measure sheds light on the technology's flexibility and long-term viability in work environments.
\end{itemize}

\subsection{Integration of Quantitative and Qualitative Data}
The results from the quantitative and qualitative analyses were integrated to provide a holistic view of the user experiences with the AR drawing system. This integration helped to validate findings and offered deeper insights into user satisfaction and areas for improvement. By combining the strengths of both quantitative and qualitative data, this analysis provides a robust foundation for understanding user interactions with the AR drawing system and informs future development and enhancements.

\section{Result}

This section presents the outcomes of our experimental evaluation involving ten participants, encompassing a diverse group of engineers and software developers. Their insights were crucial in assessing the usability and functionality of our 3D Drawing tool. Using two specific tasks designed to test the tool's capabilities, we evaluated its precision, ease of use, and the effectiveness of vectorization in a 3D annotation environment. Task 1 involved drawing a simple rock fault line, while Task 2 required outlining a complex area on a rock surface. Both tasks aimed to test the tool's ability to handle varying complexities in drawing tasks within a virtual environment.

\begin{figure}
    \centering
    \includegraphics[width=0.5\linewidth]{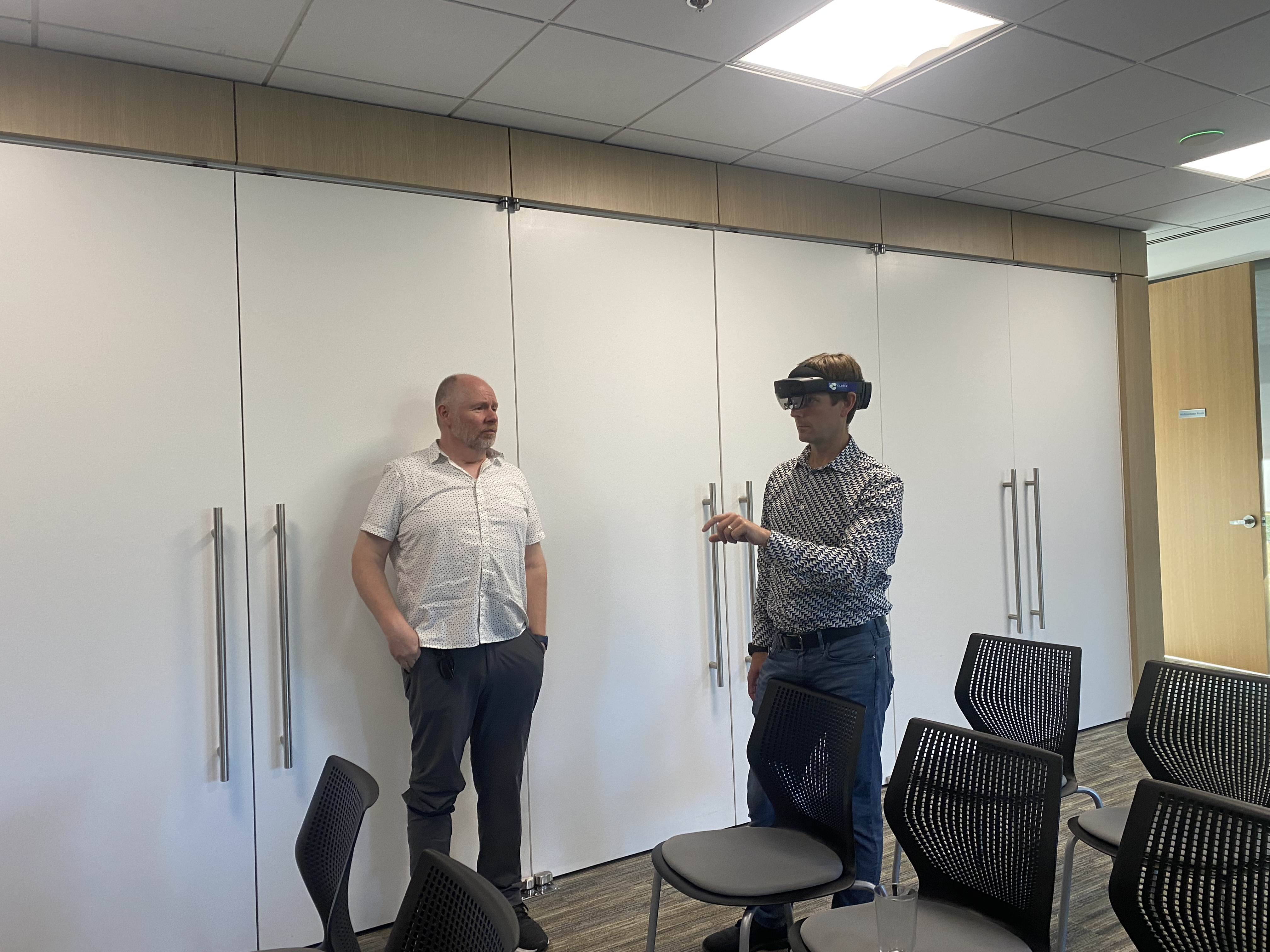}
    \caption{A participant using the 3D drawing tools}
    \label{fig:participant}
\end{figure}

\begin{enumerate}

\item \textbf{Data Size Reduction through Vectorization: }
The study assessed the impact of vectorization on data storage, revealing a significant reduction in data size across tasks. For Task 1, data size decreased by an average of 86.8\%, while Task 2 showed a reduction of approximately 85.9\%. These reductions suggest that vectorization not only conserves storage space but also potentially enhances data management efficiency in real-time applications.

\item \textbf{Impact of Vectorization on Curve Accuracy Analysis: }
Our analysis aimed to measure the deviation between curves before and after vectorization to demonstrate the process's ability to maintain precision and minimize distortion. We focused on two tasks, each involving different types of drawings within a 3D annotation environment. The data results are shown in Figure 14\ref{fig:Curve Precision Analysis}.

\textbf{Task1: }
\begin{itemize}
\item\textbf{Average Distance: }The reported average distances, such as 0.03, 0.07, and 0.1, show good adherence to the original drawings, indicating minimal loss of detail. The overall consistency suggests effective vectorization.
\item\textbf{Maximum Distance: }Maximum deviations like 0.21 highlight areas where vectorization may struggle with absolute fidelity, particularly in more complex drawing segments. However, these remain within a reasonable range.
\item\textbf{Standard Deviation: }Low standard deviations across participants indicate a reliable vectorization process, with most data showing reasonable consistency.
\end{itemize}

\begin{figure}
    \centering
    \includegraphics[width=0.8\linewidth]{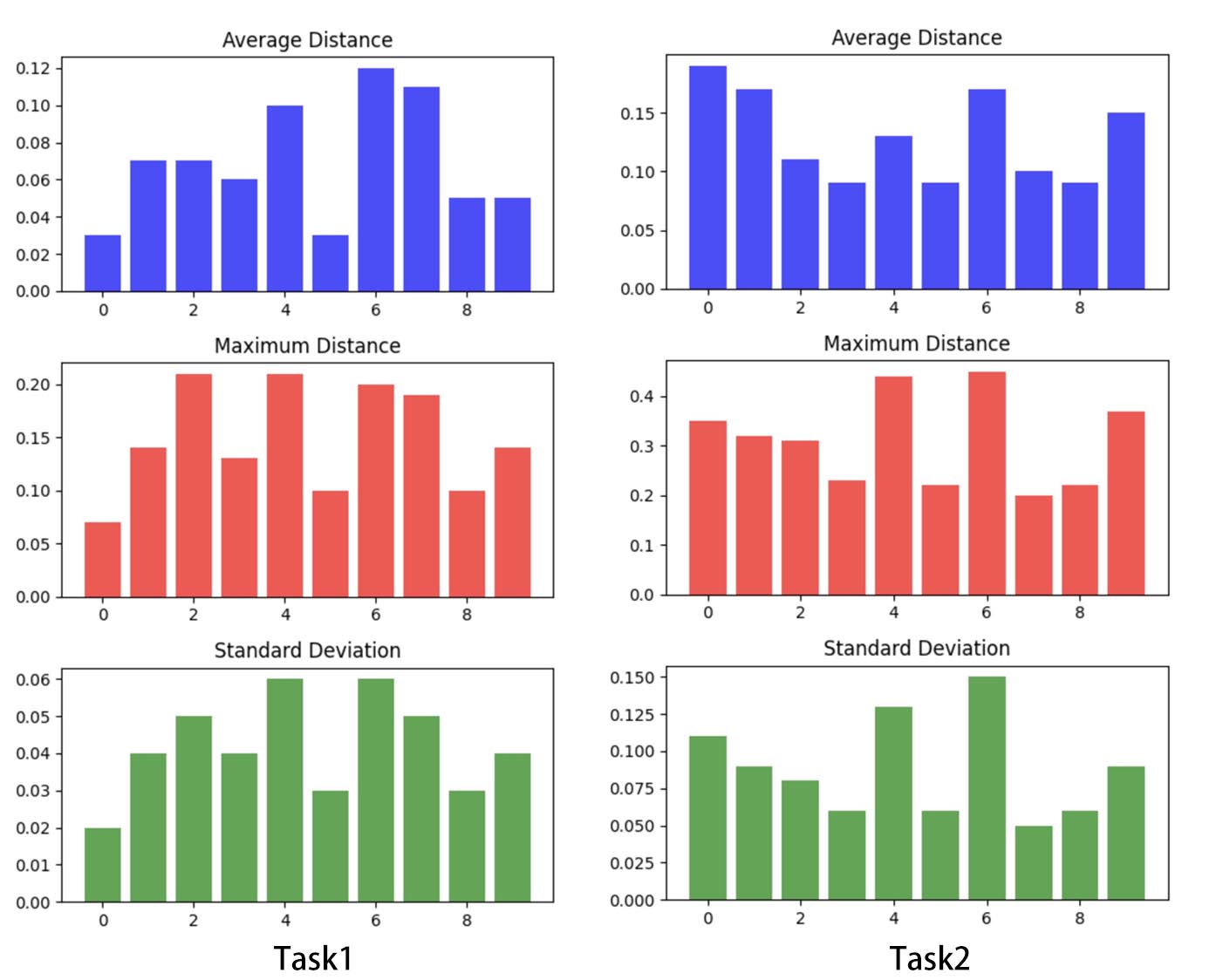}
    \caption{Impact of Vectorization on Curve Precision Analysis}
    \label{fig:Curve Precision Analysis}
\end{figure}

\textbf{Task2: }
\begin{itemize}
\item\textbf{Average Distance: }Despite the complexity of the tasks, average distances like 0.19 and 0.13 suggest that the vectorization maintains good detail.
\item\textbf{Maximum Distance: }Higher deviations, such as 0.44, indicate challenges in maintaining precision in detailed areas but are still within acceptable limits.
\item\textbf{Standard Deviation: }The variability in standard deviations shows consistency in the vectorization process, although some improvements could be beneficial.
\end{itemize}

Feedback from participants using the system supported the quantitative data. Users noted that the post-vectorization curves generally resembled the original inputs closely, which is consistent with the measured data showing that vectorization consistently maintains the overall integrity of the curve and important features.

Considering the model's dimensions (2m x 1.16m x 0.45m), we have utilized the diagonal length, approximately 2.36 meters, as a metric to evaluate deviations. This approach offers a h extensive overview on the spatial impact of these deviations within the model. An observed average deviation of about 0.1 meters represents approximately 4.24\% of the diagonal length. This percentage provides a clear indication of the deviations' relative importance throughout the model. While a 4.24\% deviation might be acceptable for general applications, it becomes significant for tasks that require high precision. This realization emphasizes the need for the vectorization technique to be further improved in order to minimize deviations, particularly in applications that call for a high degree of accuracy. Future improvements should focus on enhancing the tool's precision, particularly in managing complex shapes and detailed features.

\item \textbf{Curve Smoothness Enhancement: }

The evaluation of curve smoothness was substantiated by the analysis of average angles between vectors and the standard deviation of these angles, with the aim of illustrating the impact of vectorization on the smoothness of curve transitions. The data derived from both tasks are visualized in Figure 15\ref{fig: Enclosed Precise Feature Drawing}, showing notable differences in smoothness before and after vectorization.

\textbf{Task1: }

\begin{itemize}
\item\textbf{Average Angle Between Vectors: }The data indicates a clear reduction in the average angles between vectors post-vectorization, with figures showing a decrease across almost all data points. This suggests that vectorization helps in aligning the curve segments more smoothly, resulting in a more cohesive and continuous line.

\item\textbf{Standard Deviation of Angles: }There is a significant reduction in the variability of angles, as evidenced by the lower standard deviations after vectorization. This reduction highlights an enhanced consistency in curve smoothness, with fewer abrupt changes in direction, which is crucial for the precision required in fault line drawings.
\end{itemize}

\textbf{Task2: }
\begin{itemize}
\item\textbf{Average Angle Between Vectors: }The changes in average angles post-vectorization are less pronounced compared to Task 1, yet still present a general improvement. This indicates that while the vectorization process enhances smoothness, the complexity of the shapes in Task 2 poses challenges in achieving the same level of smoothness as seen in Task 1.
\item\textbf{Standard Deviation of Angles: }Similar to Task 1, there is a decrease in the standard deviation of angles post-vectorization, although not as substantial. This improvement shows a better consistency in the way curves are formed post-vectorization, albeit with room for further enhancement to match the smoothness seen in simpler tasks.
\end{itemize}

\begin{figure}
    \centering
    \includegraphics[width=0.8\linewidth]{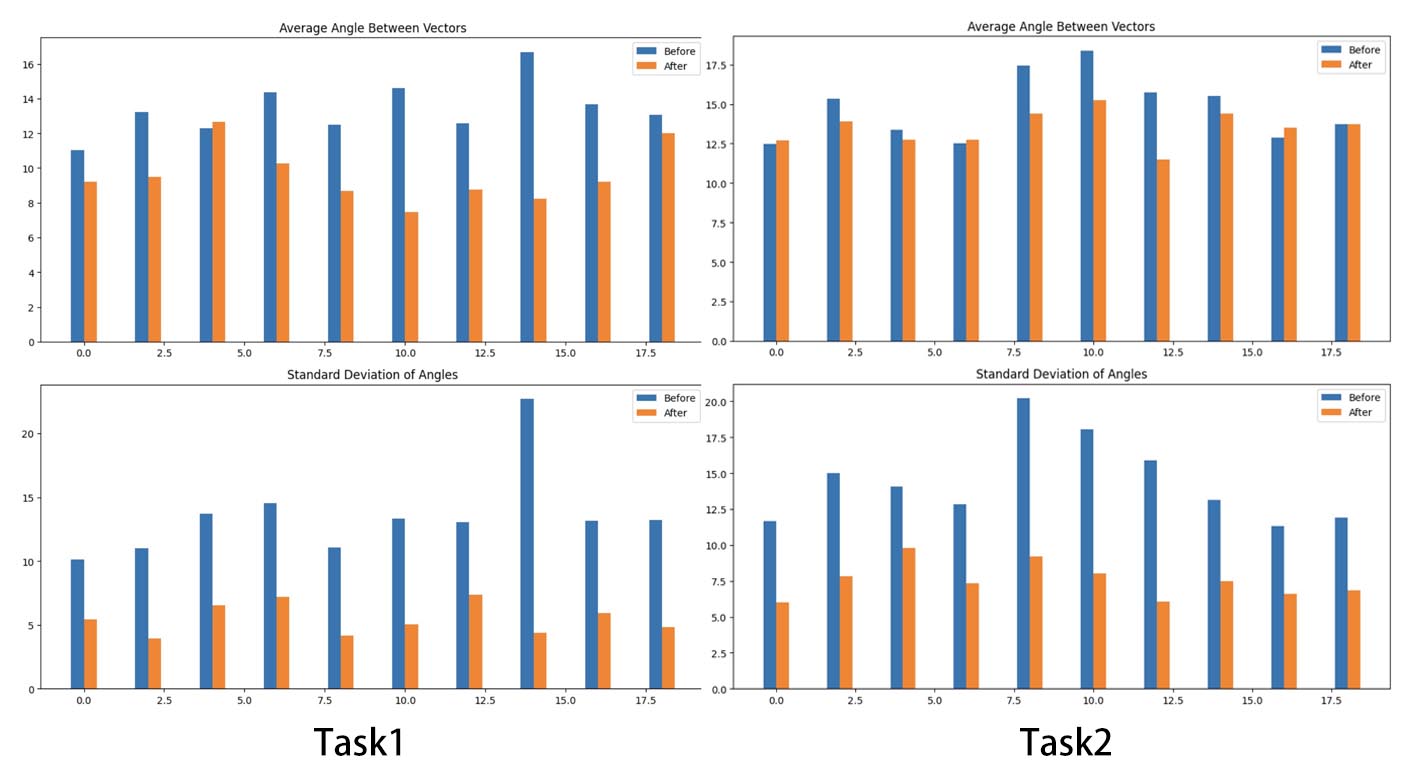}
    \caption{Curve Smoothness Enhancement}
    \label{fig:Curve Smoothness Enhancement}
\end{figure}

The appended graphs clearly depict these changes, with the 'Before' bars generally higher than the 'After' bars, especially in the context of standard deviations, corroborating the textual data with visual evidence. The visible enhancements in curve smoothness following vectorization indicate how well the vectorization process works to improve the smoothness of drawn curves.

These results highlight the positive impact of vectorization on curve smoothness across different types of tasks. Task 1 demonstrates a more significant improvement due to its less complex nature, while Task 2, despite showing progress, suggests a need for tailored vectorization strategies to handle complex geometries more effectively. Future work will focus on optimizing vectorization techniques to enhance curve smoothness further, particularly for intricate and detailed designs.

\item \textbf{Questionnaire Results: }
Questionnaire responses rated the system highly in terms of responsiveness, editing features, and detail preservation, with average scores above 4 out of 5. Participants were asked to rate their experience using the tool on three aspects on a scale from 1 (lowest) to 5 (highest): responsiveness of the tool, quality of the editing features, and how well the details were preserved in their tasks.

\begin{itemize}

\item\textbf{Responsiveness: }The tool showed excellent responsiveness, mainly receiving scores between 4 and 5. With most users reporting a responsive interface that easily satisfied their needs, the system's ability to enable rapid and efficient involvement is reflected in its high ranking.
\item\textbf{Editing Features: }Ratings for editing features were consistently high, ranging from 3 to 5, suggesting that the tool's functionality met or exceeded the expectations of most users. Although the essential functionalities are powerful, these ratings indicate that there may be space for interface improvement and the addition of features like enhanced pressure sensitivity and undo/redo capabilities.
\item\textbf{Details Preserved: }The detail preservation aspect received strong positive feedback, with average score higher than 4, showing that the tool effectively maintained the essential details of the original designs.
\end{itemize}

The uniformly high ratings across all aspects reflect a successful user interface and functionality of the tool, with particular strengths noted in responsiveness and detail preservation. The slightly lower score in editing features from one participant might indicate areas for further improvement or additional feature integration.

\item \textbf{Semi-Structured Interview Insights: }
The semi-structured interviews provided valuable qualitative data, offering insights into the user experience and potential areas for enhancement in the AR drawing system. Participants were asked about their overall experience, challenges faced, the intuitiveness of the system, and suggestions for improvements.

\textbf{Key Feedback and Suggestions: }
Ease of Use and Intuitiveness: Participants noted that the system was generally intuitive, with features like "key points to reshape" being highlighted as particularly useful. However, some mentioned that aspects such as "the feel of the cursor on the screen" could be improved for a more natural drawing experience.

\textbf{System Improvements: }
UI Enhancements: Users suggested improvements in the user interface, such as "colors need to be more " and "improved UI elements like buttons to be easier to understand". This feedback points to a need for more visual feedback and clearer interaction cues.
Feature Enhancements: Several users expressed the need for "better feedback on touch," indicating a desire for more responsive tactile feedback. Suggestions included "very likely needs a redo/undo feature" and "more responsive to pressure," which would allow for greater control over the drawing process.
Functionality: Enhancements such as "add points of the line to reshape" and "more ways to manipulate the drawing directly" were suggested to enhance the editing capabilities and flexibility of the system.

\end{enumerate}

\section{Discussion}
This study underscores the effectiveness of the tool in enabling rapid, editable, and precise 3D drawing within a mixed reality environment, as demonstrated through tasks that assessed the tool’s ability to draw and annotate complex geological features.

\textbf{Effectiveness of Vectorization: }
Our findings indicate that vectorization significantly reduces data size, with reductions of 86.8\% and 85.9\% for Tasks 1 and 2, respectively. This substantial decrease in data size not only enhances storage efficiency but also proves crucial for the processing capabilities of lightweight mobile devices. Furthermore, the vectorization process maintains a high degree of accuracy in the curves, significantly improving their smoothness and aesthetic appeal. It also increases the editability of the data, which has been highly appreciated by users, as it allows for more flexible and precise modifications.

\textbf{Challenges in Handling Complex Shapes: }
Despite the noted successes, the result in Task 2 revealed some limitations when dealing with complex shapes. Compared to Task 1, the deviations from the original curves were greater in Task 2, even if there was still some increase in the smoothness of the curves. This increased deviation highlights the need for more advanced algorithms capable of effectively managing complex details and irregular shapes. Such capabilities are particularly critical in fields like geological surveying and architecture, where precision in representing complex forms can significantly impact the outcome of practical applications. The greater deviations observed in Task 2 underscore the challenges and emphasize the necessity for algorithmic improvements to ensure high fidelity in the vectorization of complex geometries.

\textbf{User Experience and System Usability: }
Participant feedback highlighted the system's high responsiveness and editability, yet it also pointed out several areas needing improvement. Current editing features are good but limited; users expressed a desire for functionalities such as undo and redo options, an eraser tool, and the ability to add points between existing segments during the vector drawing stage. Additionally, issues were noted with button responsiveness, where touches sometimes result in delays. Enhancing these aspects would greatly improve user interaction, making the system not only more intuitive but also more accommodating to detailed user operations.


\section{Future Work}
In this project, our focus was primarily on the implementation of drawing and vectorization features. Consequently, some essential editing tools are currently not incorporated. Specifically, the addition of features such as "Undo" and "Eraser" would significantly enhance usability. However, implementing an "Undo" function requires tracking each action, which could lead to substantial performance challenges. An "Eraser" tool also presents complexities, particularly in managing data for interconnected strokes, necessitating sophisticated data handling strategies. Future enhancements should also aim to include advanced annotation capabilities and color-filling features to expand the tool’s functionality.

Feedback from survey participants has highlighted areas needing improvement, such as pressure sensitivity and gesture recognition. Notably, difficulties were reported with the Microsoft Hololens 2 in detecting pinching gestures and left-hand usage. Since this technology has been evolving since 2018, these limitations might be addressed more effectively with newer devices, like the Apple Vision Pro, suggesting a direction for future hardware integration to better support our system’s requirements.

\bibliographystyle{ACM-Reference-Format}
\bibliography{reference}

\end{document}